\documentclass[12pt]{article}
\usepackage{graphicx}

\newcommand{\bea}{\begin{eqnarray}}
\newcommand{\eea}{\end{eqnarray}}

\title{{\small\hfill IMSc/2006/10/23}\\
\textbf{Criterion for dynamical chiral symmetry breaking}}
\author{Bithika Jain$^{a}$,
Indrajit Mitra$^{b,c}$\footnote{E-mail: imitra@iitk.ac.in} and
H. S. Sharatchandra$^{d}$\footnote{E-mail: sharat@imsc.res.in} \\\\
$^a$ St. Xavier's College, Mahapalika Marg, Mumbai 400001, India\\
$^b$ Theory Group, Saha Institute of Nuclear Physics, 1/AF Bidhan-Nagar,\\
Kolkata 700064, India\\
$^c$ Department of Physics, Indian Institute of Technology,\\ Kanpur 208016, India\\
$^d$ The Institute of Mathematical Sciences, C.I.T. Campus, Taramani P.O.,\\
Chennai 600113, India}
\date{}
\begin{document}
\maketitle
\begin{abstract}

The Bethe-Salpeter equation is related to a generalized quantum-mechanical
Hamiltonian. Instability of the presumed vacuum, indicated by a tachyon, is
related to a negative energy eigenstate of this Hamiltonian. The variational method shows that
an arbitrarily weak long-range attraction leads
to chiral symmetry breaking, except in the scale-invariant case when the instability
occurs at a critical value of the coupling. In the case of short-range attraction, 
an upper bound for the critical coupling is obtained.

\end{abstract}
\newpage

%%%%%%%%%%%%%%%%%%%%%%%%%
\section{Introduction}
\label{intro}
%%%%%%%%%%%%%%%%%%%%%%%%%
There are many important contexts in which chiral or other symmetries are dynamically
broken due to a condensation of (say) fermions. It is accepted that in QCD with massless quarks,
the colour-singlet flavour-singlet quark bilinear $\bar\psi\psi$ has a non-zero vacuum
expectation value resulting in a spontaneous breaking of chiral symmetry.
Moreover, this symmetry is expected to be restored at high temperature. 
Other condensates of
fermion composites are also suggested as the baryon chemical potential is varied. 
Many
exotic condensations are suggested in technicolour theories. 
Condensation of
relativistic massless fermions is also important in the context of effective field theories for 
many condensed matter systems.

The most popular way of handling chiral symmetry breaking is to phenomenologically incorporate
it through an effective Lagrangian. Almost the only a priori calculation is
through a nonperturbative solution of the Schwinger-Dyson equation with some truncations.
This was pioneered long ago by Nambu and Jona-Lasinio \cite{njl}. 
This was in turn motivated by the gap equation
in the theory of superconductivity.

The BCS theory of superconductivity \cite{bcs} 
is founded on the Cooper pair instability.
The instability of the normal vacuum can be demonstrated for an arbitrarily small attraction
(mediated by the phonon) between the non-relativistic electrons. It may be noted that
it is easier to establish
this instability than to calculate the consequent condensation.

We would like to have a criterion for instability of the presumed vacuum for the case of
relativistic massless fermions in quantum field theory.
The analysis of a fermion-antifermion pair in this case is complicated by 
various issues in the theory of 
Bethe-Salpeter equation (BSE).
We consider Euclidean BSE at zero energy-momentum, and relate the equation to a generalized 
quantum-mechanical Hamiltonian. Instability indicated by a tachyon pole in the Bethe-Salpeter
amplitude is then related to a negative energy eigenstate
in the quantum-mechanical system.

It is to be noted that this approach does not involve solving the
gap equation. Our calculation is performed with massless fermions, and
a tachyon appears at some critical value of the coupling. This indicates instability of
the presumed, chiral symmetric  vacuum.

We now list some other works which consider the issues addressed in this paper.
The instability of the normal state in the BCS theory is demonstrated in Ref.\ \cite{schr}
using ladder diagrams for electron-electron scattering. The criterion for instability
of the chiral symmetric phase in QED corresponding to the existence of tachyonic
bound states of fermion-antifermion pair was considered in Ref.\ \cite{fomin}
(see also the review in Ref.\ \cite{nuovo}). The gap equation was related to
a quantum mechanical Hamiltonian to find the critical coupling for the onset of
chiral symmetry breaking in Ref.\ \cite{gus1} for quenched QED and in Ref.\ \cite{gus2}
for gauge theories with extra dimensions. The ladder approximation for the scalar vertex in
QED$_3$ was reduced to a quantum mechanical problem in Ref.\ \cite{herbut}.
Ref.\ \cite{khv} addresses critical couplings in
a variety of situations in condensed matter systems.    

The paper is organized as follows. In Sec.\ \ref{case}, we consider the case of massless
QED$_3$ and explain our approach for establishing the instability of the presumed vacuum.
In Sec.\ \ref{variation}, we introduce the use of the variational method for locating 
the critical coupling, and illustrate the method for massless QED$_3$. In Sec.\ \ref{massless},
the variational method is applied to treat the generic situation with massless exchange, including
both the scale-invariant and the scale-noninvariant cases. 
The case of  short range attraction is addressed in Sec.\ \ref{shortrange}.
In Sec.\ \ref{concl}, we present our conclusions. 
%%%%%%%%%%%%%%%%%%%%%%%%%%%%%%%%%%%%%%%%%%%%%%%%%%%%%%%%%%%%%%%%%%%%%%%
\section{The case of massless QED$_3$}
\label{case}
%%%%%%%%%%%%%%%%%%%%%%%%%%%%%%%%%%%%%%%%%%%%%%%%%%%%%%%%%%%%%%%%%%%%%%
To establish a proper setting, we first consider 
the case of massless QED in three (space-time) dimensions
which has many interesting features. Vacuum polarization due to $N$ massless fermion flavours in a
$1/N$ expansion changes the infrared (IR)
behaviour of the photon from inversely quadratic to inversely linear 
in momentum \cite{appel, appel'}. This feature
survives to all orders \cite{appel', mrs}. Moreover in a specific non-local gauge 
\cite{nonlocal, georgi}, 
there are no logarithmic   
IR contributions to the fermion self-energy and the vertex correction to all orders \cite{mrs}. 
Thus there is no
anomalous dimension for the fermion in this gauge, and the full vertex can be replaced by the 
bare vertex in the IR. 

%%%%%%%%%%%%%%%%%%%%%%%%%%%%%%%%%%%%%%%%%%%%%%%%%%%%%%%%%%%%%%%%%%%%%%%%%%%
%{BSE}
\begin{figure}
\begin{center}
\includegraphics[scale=1.05]{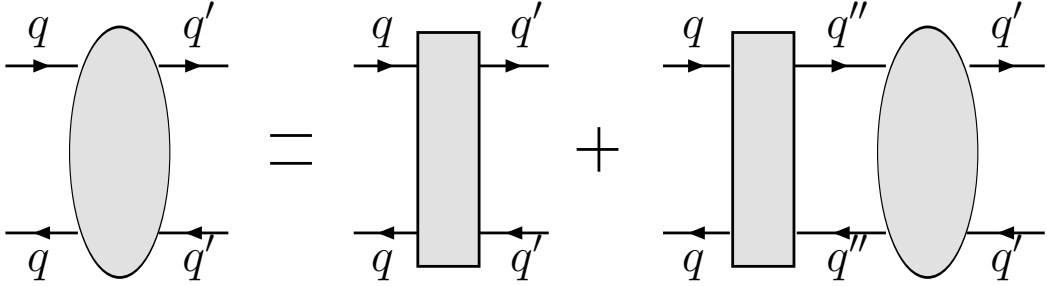}
\end{center}
\caption{The Bethe-Salpeter equation at vanishing total energy-momentum
for  the fermion-antifermion pair (the box represents the 2PI scattering kernel).}
\label{f:bethe}
\end{figure}
%%%%%%%%%%%%%%%%%%%%%%%%%%%%%%%%%%%%%%%%%%%%%%%%%%%%%%%%%%%%%%%%
%%%%%%%%%%%%%%%%%%%%%%%%%%%%%%%%%%%%%%%%%%%%%%%%%%%%%%%%%%%%%%%%
%{Scalar channel}
\begin{figure}
\begin{center}
\includegraphics[scale=0.9]{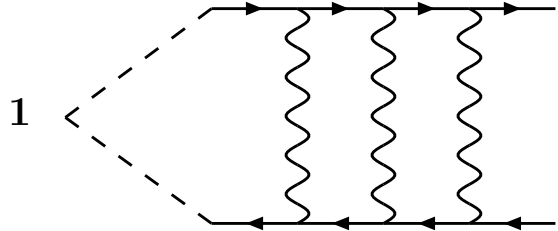}
\end{center}
\caption{The scalar channel leads to a drastic simplification in the spinor algebra.}
\label{f:scalar}
\end{figure}
%%%%%%%%%%%%%%%%%%%%%%%%%%%%%%%%%%%%%%%%%%%%%%%%%%%%%%%%%%%%%%%%%%%%%%%%%%%%

Therefore we consider the BSE in the ladder approximation: this is 
given by Fig.\ \ref{f:bethe}
with the simplest kernel in the form of a single photon exchange. 
Let us now consider the following special situation \cite{mrs'}.
We work with the Euclidean theory. The photon propagator is replaced with its IR behaviour, 
which suffices because
 any instability is expected to be driven by the attraction at small 
momentum transfer. We consider 
fermion-antifermion scattering at vanishing total energy-momentum and in the
scalar channel (see Fig.\ \ref{f:scalar}). This provides a drastic simplification in the
spinor algebra, and we are rid of all the spinor and Lorentz indices. We then obtain
a simple integral equation: 
\bea
I(\vec q, \vec q\,')=\frac{1}{|\vec q-\vec q\,'|} +\lambda\int d^3 q''\frac{1}{q''^2}
                   \frac{I(\vec q\,'', \vec q\,')}{|\vec q- \vec q\,''|}\,.    
\eea
Here the coupling $\lambda$ equals $8/3\pi^3 N$ to the lowest order in $1/N$ \cite{mrs'}. 
We may convert this integral equation into a differential equation by applying $\nabla{_q^2}$:
\bea
\Bigg(-\nabla{_q^2}-\frac{4\pi\lambda}{q^2}\Bigg)I(\vec q,\vec q\,')
                     =4\pi\delta^{(3)}(\vec q-\vec q\,')\,.         
\eea
Thus $I(\vec q, \vec q\,')$ is the Green function for the Hamiltonian of a three-dimensional Schr\"odinger equation
(in variables $\vec q$) with an attractive inverse square potential
\bea
V(q)=-4\pi\lambda/{\vec q}\,^2\,.                                     \label{pot}
\eea
Therefore
\bea
I(\vec q, \vec q\,')=4\pi\sum_n \frac{\chi_n(\vec q)\chi{^*_n}(\vec q\,')}{\epsilon_n}\,,    \label{green}
\eea
where $\chi_n(\vec q)$ are the eigenfunctions of the Hamiltonian with eigenvalues $\epsilon_n$. This is a well-studied
problem in quantum mechanics with unusual properties \cite{landau}. When the attraction is smaller than a critical value,
$\lambda\le\lambda_c=1/16\pi$, the Hamiltonian has a continuous spectrum starting at $\epsilon=0$. For
$\lambda > \lambda_c$, the spectrum is unbounded from below.

As we are considering the Euclidean BSE with a vanishing total incoming momentum, $\epsilon_n < 0$ in Eq.\ (\ref{green})
corresponds to {\it a tachyonic bound state of the fermion-antifermion pair}. This signals
instability of the system
against chiral symmetry breaking \cite{mrs'}.
A study of the gap equation by other authors has led to a dynamical mass formation at precisely this critical 
value of the coupling \cite{nonlocal}. 
The important thing to note is that {\it the approach simplifies 
the problem 
in so much as relating it to a negative energy eigenstate for a quantum mechanical Hamiltonian.}
This simplification is utilized below.

In this model, there is a sudden change to arbitrarily negative energy values at a $\lambda_c$.
This is an artifact of this particular model: exact scale-invariance (in the IR)
leads to a continuum of negative
energy eigenvalues, once there is one such eigenvalue.
 
Note also that the inverse square potential is coming 
from the massless fermion, and not from the forces of attraction.
The photon propagator gives rise to the `kinetic energy' operator 
($-\nabla{_q^2}$ being the operator
for which the propagator $1/|\vec q-\vec q\,'|$ is the Green function).

%%%%%%%%%%%%%%%%%%%%%%%%%%%%%%%%%%%%%%%%%%%%%%%%%%%%%%%%%%%%%%%%%%%%%%%%%%%
\section{The variational method}
\label{variation}
%%%%%%%%%%%%%%%%%%%%%%%%%%%%%%%%%%%%%%%%%%%%%%%%%%%%%%%%%%%%%%%%%%%%%%%%%%%
In the case discussed in Sec.\ \ref{case}, we were 
fortunate in having an exactly solvable quantum-mechanical Hamiltonian to locate the
critical value of the coupling. This is not the situation in the other cases to be discussed below.
We apply the variational method to locate the critical value of the coupling at which a tachyon is formed.
We first illustrate the technique for the case of massless QED$_3$.

We want an estimate of the ground state energy. 
For this we should use a trial wave function which is large
at $q=0$ so that the potential energy of Eq.\ (\ref{pot}) has an expectation value 
$\langle V\rangle$  (which is necessarily 
negative definite) as large in magnitude as possible. In addition, 
the wave function should not fall off
too rapidly in order that the the expectation value $\langle T\rangle$ of the kinetic energy (necessarily
positive definite) is not too large. 
Let us therefore use a trial state
\bea
\psi(\vec q)=q^{-s}e^{-\mu q}                              \label{trial}
\eea
with $s>0$ and $\mu>0$, using which we can make exact calculations.  
Normalizability of this wave function requires $3-2s>0$. Requiring that $\langle V\rangle$ is finite gives a stronger
condition $1-2s>0$. In that case, $\langle T\rangle$ is also finite. We have
\bea
||\psi||^2=4\pi(2\mu)^{2s-3}\Gamma(3-2s)\,,                                                  \label{eq:norm}
\eea
\bea
\langle\psi|V|\psi\rangle=-16\pi^2\lambda(2\mu)^{2s-1}\Gamma(1-2s)\,.                         \label{eq:T}
\eea
As the wave-function has no angular dependence, $\nabla^2 \psi$ is effectively $(1/q)d^2(q\psi)/dq^2$ and 
\bea
\langle\psi|(-\nabla^2)|\psi\rangle=2\pi(2\mu)^{2s-1}(1-s)\Gamma(1-2s)\,.                        \label{eq:V}     
\eea
Adding Eqs.\ (\ref{eq:T}) and (\ref{eq:V}) and dividing by $||\psi||^2$, we obtain
\bea
\langle E\rangle=\frac{\mu^2}{1-2s}\Bigg[1-\frac{8\pi\lambda}{1-s}\Bigg]\,.                   \label{eq:E}
\eea
Since $1-2s>0$, and therefore, $1-s>1/2$, the quantity within the square brackets cannot be made negative for
$\lambda<1/16\pi$. Whereas for $\lambda>1/16\pi$, one can increase $\mu$ to get arbitrarily negative value
for $\langle E\rangle$. We thus get the correct value for the critical coupling. The change of the minimum possible value of 
$\langle E\rangle$ from zero to $-\infty$ as $\lambda$ crosses $\lambda_c$ is a consequence of the scale
invariance in the problem.

%%%%%%%%%%%%%%%%%%%%%%%%%%%%%%%%%%%%%%%%%%%%%%%%%%%%%%%%%%%%%%%%
\section{Generic situation with massless exchange}
\label{massless}
%%%%%%%%%%%%%%%%%%%%%%%%%%%%%%%%%%%%%%%%%%%%%%%%%%%%%%%%%%%%%%%%
We now apply our technique to a generic situation with massless exchange (Fig.\ \ref{f:bethe}). 
We work with the renormalized theory. Presuming the chiral symmetric vacuum, we consider 
the IR behaviour
{\it after all IR divergences in this theory have been taken into account.}
Let the leading behaviour of the
2PI fermion-antifermion scattering kernel in the scalar channel for small momentum transfer
$|\vec q-\vec q\,'|\rightarrow 0$ be $\kappa/|\vec q-\vec q\,'|^\alpha$. (As earlier, we consider the
case of zero total incoming momentum.) We also allow an anomalous dimension for the massless fermions,
and take the fermion propagator to be $\rlap q//q^{\beta+1}$ in the IR. Thus the BSE (keeping only the
leading IR behaviour) in $d$ space-time dimensions is
\bea
I(\vec q, \vec q\,')=\frac{1}{|\vec q-\vec q\,'|^\alpha} +\lambda\int d^d q''\frac{1}{q''^{2\beta}}
                   \frac{1}{|\vec q- \vec q\,''|^\alpha}I(\vec q\,'', \vec q\,')\,.                  \label{eq:bse}  
\eea
Here $\lambda=\kappa/(2\pi)^d$, and we have factored out 
one factor of $\kappa$ from each term of the equation.
We now obtain the formal operator for which the Green function is $1/|\vec q-\vec q\,'|^\alpha$. Consider the
Fourier transform
\bea
\frac{1}{|\vec q- \vec q\,'|^\alpha}=c(d,\alpha)\int\frac{d^dp}{(2\pi)^d}e^{i\vec p\cdot(\vec q-\vec q\,')}
                                    \frac{1}{|\vec p|^{d-\alpha}}\,.
\eea
The required operator is therefore $(1/c(d,\alpha))(\vec {\cal P}^2)^{(d-\alpha)/2}$ where
${\cal P}_\mu=-id/dq_\mu$. We
may regard $\vec q$ as the `coordinate' variables and $\vec p$ as the conjugate `momentum' variables,
with the corresponding operators $\vec{\cal Q}$ and $\vec{\cal P}$ (with $\hbar=1$). 
We can now formally relate the integral equation (\ref{eq:bse}) to a generalized quantum-mechanical Hamiltonian:
\bea
H=(\vec{\cal P}^2)^{(d-\alpha)/2}-\lambda\, c(d,\alpha) \frac{1}{(\vec{\cal Q}^2)^\beta}\,,
\eea
\bea
HI(\vec q, \vec q\,')=c(d,\alpha)\delta^{(d)}(\vec q- \vec q\,')\,.
\eea 

As in the previous case, we want to 
know when there is a negative eigenvalue for this Hamiltonian. It suffices to
choose the same trial wave-function as given in Eq.\ (\ref{trial}). The calculation of $\langle T\rangle$
is easier when $d$ is an odd integer (because the angular integration is simple). Therefore we 
present the calculations
only for $d=3$. 
To calculate $\langle\psi| T|\psi\rangle$, we use the wave-function $\tilde\psi(\vec p)$ in
the `momentum' variables, since $T$ is diagonal in the momentum basis:
\bea
\tilde\psi(\vec p)&=&\int d^3q\,e^{i\vec p\cdot\vec q}\psi(\vec q)\nonumber\\
                  &=&\frac{2\pi}{ip}\Gamma(2-s)\Bigg(\frac{1}{(\mu-ip)^{2-s}}-
                     \frac{1}{(\mu+ip)^{2-s}}\Bigg)\,.
\eea
We then obtain
\bea
\langle\psi| (\vec{\cal P}^2)^{(3-\alpha)/2}|\psi\rangle&=&\int \frac{d^3p}{(2\pi)^3}
             |\tilde\psi(\vec p)|^2 p^{3-\alpha}\\
             &=&2[\Gamma(2-s)]^2 \mu^{2s-\alpha}\nonumber\\
                &&\times\Bigg[B\Bigg(2-\frac{\alpha}{2},\frac{\alpha}{2}-s\Bigg)
                 -2\cos\Bigg(\Big(2-\frac{\alpha}{2}\Big)\pi\Bigg)B(4-\alpha,\alpha-2s)\Bigg]
\label{eq:TT}
\eea
where we put $p=\mu\tan\theta$ when doing the radial integration and used the integrals 
given in Eqs.\ (\ref{formula1}) and (\ref{formula2}).
Also,
\bea
\langle\psi| V|\psi\rangle=-4\pi\lambda\, c(3,\alpha)(2\mu)^{2s+2\beta-3}\Gamma(3-2s-2\beta)\,.
\label{eq:VV}
\eea
Adding Eq.\ (\ref{eq:TT}) and Eq.\ (\ref{eq:VV}) and 
dividing by $||\psi||^2$ (as given by Eq.\ (\ref{eq:norm})), 
we obtain
$\langle E\rangle$ for the general case in three dimensions.

{\it We first consider the scale-invariant case.} As a first example, 
it can be checked that for the QED$_3$ case $\alpha=1$ and $\beta=1$, the
above general result for $\langle E\rangle$ correctly reproduces
Eq.\ (\ref{eq:E}). (The formula given in Eq.\ (\ref{formula3}) is useful for this purpose.
Note also that $c(3,1)=4\pi$.) 
Next consider the case $\alpha=2$, $\beta=1/2$ in three dimensions. 
This case is also scale-invariant.
Now we have to use $c(3,2)=2\pi^2$. We find
\bea
\langle E\rangle=\frac{2\pi^2\mu}{1-s}\Bigg[\frac{1}{2\pi^{5/2}}\frac{\Gamma(3-s)}
{\Gamma(5/2-s)}-\lambda\Bigg]\,.
\eea
Here $s<1$ from finiteness of $\langle\psi| T|\psi\rangle$ and $\langle\psi| V|\psi\rangle$.
The smallest value of $\Gamma(3-s)/\Gamma(5/2-s)$ for $0<s<1$ is $2/\sqrt\pi$ for $s=1$.
This gives $\lambda_c=1/\pi^3$ using similar argument as presented after Eq.\ (\ref{eq:E}).

{\it Let us now consider the general scale-invariant case}
 $d-\alpha=2\beta$ in $d$ space-time dimensions, so that $T$ and $V$ scale in 
the same way. By scaling $q\rightarrow q/\mu$, we immediately arrive at the form 
\bea
\langle E\rangle=\mu^{2\beta}f_1(s)[1-\lambda/f_2(s)]\,.
\eea
Here $f_1(s)$ and $f_2(s)$ are positive definite functions of $s$ in the
allowed range of values of $s$. (This is because $\langle T\rangle$ is positive definite and 
$\langle V\rangle$ is negative definite.) Following the earlier arguments, we conclude that
the minimum value of 
$f_2(s)$ gives the critical coupling $\lambda_c$.
Again $\langle E\rangle$ can be made arbitrarily negative by taking $\mu$ to be arbitrarily large when
$\lambda > \lambda_c$.

{\it We next consider the scale-noninvariant case} $d-\alpha\neq 2\beta$. Consider first, 
for example, the case $\alpha=1$, $\beta=1/2$ in three dimensions.
In this case the general formula gives
\bea
\langle E\rangle=\frac{\mu^{2}}{1-2s}-\frac{4\pi\lambda\mu}{1-s}\,.     \label{eq:20}
\eea
For any given $\lambda$, one can choose $\mu<4\pi\lambda(1-2s)/(1-s)$ with 
any $s$ in the allowed range $0<s<1/2$ 
to have a negative value for $\langle E\rangle$.
Thus there is instability  for any coupling. (Note that on completing the square in 
Eq.\ (\ref{eq:20}), one finds that the minimum possible value of $\langle E\rangle$
is $-4\pi^2\lambda^2$, which is finite as expected in the absence of scale invariance.)

{\it For the general scale-noninvariant case}
in $d$ dimensions, 
\bea
\langle E\rangle=g_1(s)\mu^a-\lambda g_2(s)\mu^b  
\eea
where $g_1(s)$ and $g_2(s)$ are positive definite functions of $s$, and $a\neq b$. For $a>b$, one can choose  
any $s$ in the allowed range and $\mu<(\lambda g_2(s)/g_1(s))^{1/(a-b)}$ to have 
a negative $\langle E\rangle$ for any given $\lambda$.
Similarly for $b>a$, one only needs to chose $\mu$ large enough.
Thus, {\it chiral symmetry breaking takes place for an arbitrarily weak attraction
in the scale-noninvariant case involving massless exchange.}
It may be noted that this is the situation for QCD in four space-time dimensions.

%%%%%%%%%%%%%%%%%%%%%%%%%%%%%%%%%%%%%%%%%%%%%%%%%%%%%%%%%%%%%%%%%%%%
\section{The case of short-range attraction}
\label{shortrange}
%%%%%%%%%%%%%%%%%%%%%%%%%%%%%%%%%%%%%%%%%%%%%%%%%%%%%%%%%%%%%%%%%%%%%
Finally, let us address the consequences of a short-range attraction. 
To present the main features, we limit ourselves to a massless fermion of canonical dimension
and an attractive force provided by the free propagator of the mediator. Now the BSE is
\bea
I(\vec q, \vec q\,')=\frac{1}{|\vec q-\vec q\,'|^2+m^2} +\lambda\int d^3 q''\frac{1}{q''^{2}}
                   \frac{1}{|\vec q- \vec q\,''|^2+m^2}I(\vec q\,'', \vec q\,')\,.                  \label{eq:bsemass}  
\eea
Since
\bea
\frac{1}{|\vec q-\vec q\,'|^2+m^2}=2\pi^2\int\frac{d^3p}{(2\pi)^3}\,e^{-i\vec p\cdot(\vec q-\vec q\,')}
                \frac{e^{-mp}}{p}\,,
\eea
the formal operator for which this is the Green function is given by
\bea
T\sim\sqrt{\vec{\cal P}^2}e^{m\sqrt{\vec{\cal P}^2}}
\eea
Note that the spectrum blows up exponentially.

We cannot take the earlier trial wave-function because $\langle\psi| T|\psi\rangle$ does not exist:
$\tilde\psi(p)$ falls off only as power of $p$ for large $p$. $T$ is not an operator on the Hilbert
space of such wave-functions. We need to choose a different trial wave-function precisely because 
the spectrum of $T$ blows up so fast. Indeed we cannot choose wave-functions which are singular near $q=0$,
as we did earlier to gain as much potential energy as possible. 
Any power law behaviour of $\psi(q)$ as $q\rightarrow 0$
translates to a power law behaviour of $\tilde\psi(p)$ as 
$p\rightarrow \infty$, and then $\langle\psi| T|\psi\rangle$
will not be finite. This is the reason that it is so difficult to have negative energy
eigenvalues now. We choose the simple
trial wave-function 
\bea
\tilde\psi(p)=\frac{e^{-\mu p}}{p}
\eea
which corresponds to
\bea
\psi(q)=\frac{1}{2\pi^2(q^2+\mu^2)}\,,
\eea
so that we can make explicit computations.
Thus $\langle\psi| T|\psi\rangle$ is finite if $\mu>m/2$ and we obtain
$||\psi||^2=1/(4\pi^2\mu)$,
$\langle\psi| T|\psi\rangle= 1/(2\pi^2(2\mu-m)^2)$ and
$\langle\psi| V|\psi\rangle= -\lambda/(2\mu^3)$.
Therefore
\bea
\langle E \rangle=\frac{\mu^3-4\pi^2\lambda (\mu-m/2)^2}{2\mu^2(\mu-m/2)^2}\,.
\eea
Now $f(\mu)\equiv\mu^3-4\pi^2\lambda (\mu-m/2)^2$ is positive both at $\mu=m/2$ and for large $\mu$.
It is negative for a range of values of $\mu$ in between, provided $\lambda$ is sufficiently large;
then $f(\mu)$ possesses a minimum. This minimum value of $f(\mu)$ is precisely zero when $\lambda=\lambda_c$.
Therefore setting both $f(\mu)$ and $df/d\mu$ equal to zero for $\lambda=\lambda_c$, we obtain 
$\lambda_c=27m/32\pi^2$. Since other choices of $\psi(q)$ may give a lower value for
$\lambda_c$, this is actually an upper bound
for the critical coupling.

%%%%%%%%%%%%%%%%%%%%%%%%%%%%%%%%%%%%%%%%%%%%%%%%%%%%%%%%%%%%%%%%%%%%%%%%%%%
\section{Conclusion} 
\label{concl}
%%%%%%%%%%%%%%%%%%%%%%%%%%%%%%%%%%%%%%%%%%%%%%%%%%%%%%%%%%%%%%%%%%%%%%%%%
In this paper, we presented a criterion for instability of the vacuum against
chiral symmetry breaking. The instability, as indicated by a tachyon pole
in the Bethe-Salpeter amplitude, was related to a negative energy eigenstate of a
quantum-mechanical Hamiltonian. Our approach was illustrated for chiral symmetry breaking
in massless QED$_3$, and the variational method was then applied to address the critical
coupling for generic situations. 
%%%%%%%%%%%%%%%%%%%%%%%%%%%%%%%%%%%%%%%%%%%%%%%%%%%%%%%%%%%%%%%%%%%%%%%%%
\section*{Acknowledgements}
%%%%%%%%%%%%%%%%%%%%%%%%%%%%%%%%%%%%%%%%%%%%%%%%%%%%%%%%%%%%%%%%%%%%%%%%%
B.J. thanks IMSc, Chennai for a summer studentship during which this work was
initiated. I.M. thanks IMSc, Chennai for hospitality during the
course of this work. I.M. also thanks R. Jackiw for comments, and
V. Gusynin, D. Khveshchenko and B. H. Seradjeh for communications. 
The LaTeX codes for the figures in this paper were generated primarily using
JaxoDraw \cite{jaxo}.

%%%%%%%%%%%%%%%%%%%%
\appendix
\leftline{\null\hrulefill\null}\nopagebreak
\section*{Appendix}
%%%%%%%%%%%%%%%%%%%%%%%
Some of the integrals and formulas used in this paper are listed below \cite{grad}.
\bea
\int_0^{\pi/2}dx\,\sin^{m-1}x\, \cos^{n-1}x=\frac{1}{2}B\Bigg(\frac{m}{2}, \frac{n}{2}\Bigg)
\label{formula1}
\eea

\bea
\int_0^{\pi/2}dx\,\sin^{m-1}x \,\cos^{n-1}x\,\cos(m+n)x=\cos\frac{m\pi}{2}B(m,n)
\label{formula2}
\eea

\bea
B(m,n)=\Gamma(m)\Gamma(n)/\Gamma(m+n)
\eea

\bea
\Gamma(2m)=\frac{2^{2m-1}}{\sqrt\pi} \Gamma(m)\Gamma\Bigg(m+\frac{1}{2}\Bigg)
\label{formula3}
\eea

\end{document}